\documentclass[natbib]{svjour3}                     
\smartqed  
\usepackage{graphicx}
 \usepackage{aps-bibstyle}  
\newcommand{\aap}{{Astron. Astrophys.}}
\newcommand{\apj}{{Astrophys. J.}}
\newcommand{\grl}{{Geophys. Res. Lett.}}

\newcommand{\mnras}{{Mon. Not. Roy. Astron. Soc.}}
\newcommand{\jgr}{J. Geophys. Res.}
\def\mbf#1{\mbox{\boldmath ${#1}$}}
\def\Alfven{Alfv\'{e}n~}
\def\Alfvenic{Alfv\'{e}nic~}
\def\lesssim{\; \buildrel < \over \sim \;}
\def\gtrsim{\; \buildrel > \over \sim \;}

\begin{document}

\title{Self-consistent Simulations of \Alfven Wave Driven Winds  
}
\subtitle{from the Sun and Stars}


\author{Takeru K. Suzuki 
}


\institute{Takeru K. Suzuki \at
              Furo-cho, Chikusa, Nagoya, 464-8601, Japan \\
              Tel.: +81-52-788-6196\\
              Fax: +81-52-788-6196\\
              \email{stakeru@nagoya-u.jp}           
}

\date{Received: date / Accepted: date}

\maketitle

\begin{abstract}
We review our recent results of \Alfven wave-driven winds. 
First, we present the result of a self-consistent 1D MHD simulations 
for solar winds from the photosphere to interplanetary region. 
Here, we emphasize the importance of the reflection of \Alfven waves 
in the density stratified corona and solar winds.  We also introduce 
the recent HINODE observation that might detect the reflection signature 
of transverse (\Alfvenic) waves by Fujimura \& Tsuneta (2009). 
Then, we show the results of \Alfven wave-driven winds from red giant stars. 
We explain the change of the atmosphere properties from steady coronal 
winds to intermittent chromospheric winds and discuss how the wave reflection 
is affected by the decrease of the surface gravity with stellar evolution.  
We also discuss similarities and differences of accretion disk winds 
by MHD turbulence. 

\end{abstract}

\section{Introduction}
\label{intro}


The main difficulty of the coronal heating and solar wind acceleration 
is how to lift up nonthermal (e.g. magnetic) energy from the photosphere 
to upper regions because the hotter corona cannot stably exist above the 
cooler photosphere by upward thermal heat flux. In addition, 
it is not well understood how to let the energy dissipate at appropriate 
positions.   
The origin of energy that heats up the corona and accelerates the solar winds
is in the surface convective layer. The solar atmosphere is filled with 
complicated structure of magnetic field which is supposed to be amplified 
by dynamo mechanism in the interior \citep[e.g.,][]{bru04}. 
The turbulent motions of the surface convection drive various modes of 
upward propagating waves. The turbulent motions may also trigger  
magnetic reconnections which result in flares and flare-like events, 
which drive various mode of waves at locations above the 
photosphere \citep{str99}. 
In this paper, we focus on the roles of the waves in heating and accelerating 
the solar winds.

The compressive waves that are excited at the photosphere cannot travel to 
sufficiently upper regions because 
they easily steepen the wave fronts to form shocks after the amplification 
of the amplitude in the density decreasing atmosphere. For instance, 
the acoustic waves mostly dissipate before reaching the corona, and therefore 
they cannot contribute to the heating of the corona and the acceleration of
the solar wind \citep{ss72,suz02}. 
Fast mode waves in low $\beta$ (magnetically dominated) condition suffer 
refraction so that they hardly reach the coronal height if the fast mode 
speed varies in horizontal direction due to complicated magnetic structure 
\citep{kc06}. 
In general, the compressive waves 
that are excited from the photosphere cannot give 
a significant contribution to the coronal heating and solar wind acceleration. 

On the other hand, \Alfven waves can travel a longer distance owing
to the incompressive characters. A fraction of the \Alfven waves excited 
at the photospheric level is supposed to propagate 
to the corona and the solar wind acceleration region so that they
play a more dominant role in the heating and acceleration of the solar wind 
than the compressive waves. 
\Alfvenic oscillations are also detected in various regions on the Sun by 
the recent observations using the HINODE satellite,  
{\it e.g.}, in a solar prominence 
\citep{oka07}, in spicules \citep{dep07}, and in a chromospheric jet 
\citep{nis08}.

To understand the dissipation mechanism of the \Alfven waves is a key to 
understand the heating and acceleration 
of the solar wind because this corresponds to the transfer of the energy 
flux of the \Alfven waves to the ambient plasma. 
The \Alfven speed in general largely changes with height because the 
density and magnetic field strength decrease in a different manner.
\citet{shi10} have estimated the distribution of \Alfven speeds in a
polar region from HINODE observation and found that the \Alfven speeds 
actually vary in both vertical and horizontal directions. 
Under these circumstances, \Alfven waves suffer reflection through the 
upward propagation as a result of the deformation of the wave shape 
\citep{an90,moo91,hi07}. 
Interactions between the outgoing \Alfven waves and the incoming component  
possibly lead to the damping through turbulent-like cascade \citep[e.g.,][]
{dmi03,cra07,vv07}. 
This process might further generate high-frequency 
ioncyclotron waves, which are widely discussed especially on the preferential 
heating of the perpendicular components with respect to magnetic field of 
minor heavy ions \citep[e.g.,][]{tm97,koh98,cra99}.
The variation of the \Alfven speeds also anticipate phase mixing of \Alfven 
waves along neighboring field lines \citep{hp83,sg84,vg05}, which might 
further generate fast mode waves \citep{nak97}.  
The density stratification and the variation of the \Alfven speed also 
enhances the parametric decay instability of outgoing \Alfven waves 
(Suzuki \& Inutsuka 2006; SI06 hereafter). 
The parametric decay excites compressive waves in addition 
to incoming \Alfven waves.    
In contrast to those excited at the photosphere, these compressive waves 
are generated in the upper regions and directly contribute to the heating 
of the corona and solar winds by the shock dissipation.

Although the final wave dissipation mechanism and heating process 
are still not well figured out, understanding the wave reflection in 
stratified atmosphere with varying \Alfven speed is primarily important. 
\Alfven waves are reflected most effectively in the 
chromosphere and transition region because of the rapid change of the 
density. 
While most of the \Alfven waves from the surface cannot penetrate into 
the corona, the remaining waves sufficiently heat and accelerate the solar 
wind. Our previous work (Suzuki \& Inutsuka 2005; SI05 hereafter) showed 
that the 
transmitted fraction into the corona is an order of 10$\%$ and that this is 
enough for the coronal heating and solar wind acceleration in open coronal 
holes. 
Although turbulent-like cascade of \Alfven(ic) waves is not treated in this 
work because we adopted one dimensional (1D) magnetohydrodynamical (MHD) 
simulations, this is the first attempt that dynamically treats wave reflection 
from the photosphere to solar wind regions simultaneously with heating of 
the gas by the wave damping. In this review paper, we firstly summarize 
our dynamical simulations of solar winds, focusing on the reflection 
of \Alfven waves under various conditions.  
Actually, the recent HINODE observation also detected a signature of 
reflected \Alfven waves \citep{ft09}, which will be introduced in more 
detail later. 

The concept of \Alfven wave-driven winds is not limited
to the Sun. It is expected that the similar mechanism operates in 
winds from other stars \citep[e.g.][]{vel93}, such as protostars \citep{cra08}, 
low- and intermediate-mass main sequence stars, red giant stars 
\citep[][S07 hereafter]{suz07}, and proto-neutron stars with strong magnetic 
field (Suzuki \& Nagataki 2005). 
Among these objects, we show \Alfven wave driven 
red giant winds in this paper based on our work (S07). 
In red giant stars, the stratification properties (e.g. decrease of density) 
of the atmosphere are different from those in the Sun because of the smaller 
surface gravity. This affects the propagation and reflection of 
the \Alfven waves. Here we discuss the change of the wave reflection with 
stellar evolution.

In the last part of this paper, we further extend to winds from accretion 
disks. To date, it has been widely discussed that accretion disk winds 
can be driven by centrifugal force \citep{bp82,ks98}. 
Global magnetic field plays a key role in this process; if there is poloidal 
field that is sufficiently tilted with respect to an accretion disk, 
the gas can flow out along with the field lines.
On the other hand, it is expected that small-scale MHD turbulence also 
potentially drives disk winds, similarly to the solar winds, 
although such mechanism has not been well studied so far. 
It is now widely accepted that the outward transport of 
disk angular momentum and inward mass accretion are realized by the 
effective viscosity owing to turbulence. Magnetorotational instability (MRI)
is now known to be a very effective mechanism that drives turbulence if weak 
magnetic fields exist initially (Balbus \& Hawley 1991). 
The excited MHD turbulence is supposed to accelerate the disk material 
to upper regions as well as transport the angular momentum outwardly. 
In Suzuki \& Inutsuka (2009), we studied the disk winds driven by MHD 
turbulence by local disk simulations. In this paper, we introduce the 
results with emphasizing the similarity to and difference from the \Alfven 
wave-driven solar and stellar winds.

\section{Solar Winds}
\label{sec:slw}
\subsection{Simulation Set-up}

\begin{figure}
  \includegraphics[width=0.8\textwidth]{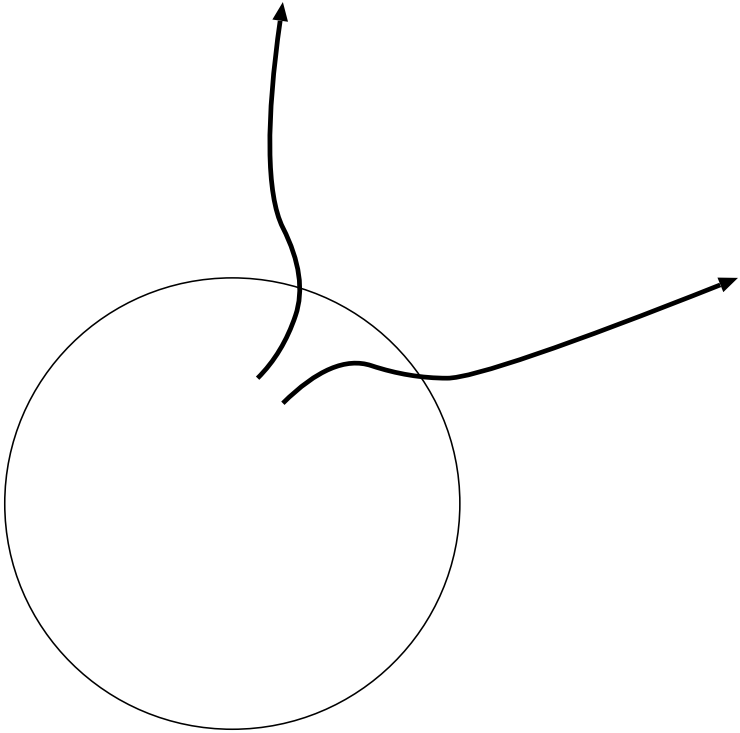}
\caption{Set-up of solar wind simulations. The cartoon shows 
a schematic view of a super-radially open flux tube, which we set 
in our simulations.}
\label{fig:scm}       
\end{figure}

We simulate the heating and acceleration of solar winds in 1D super-radially 
expansing flux tubes from the photosphere to sufficiently distant locations 
(0.1 -- 0.3 AU). Here we summarize the basic points of the simulation set-up. 
For the details, please see SI05 and SI06. 
We determine radial magnetic field strength, $B_r$, from the conservation 
of magnetic flux : 
\begin{equation}
B_r f r^2= {\rm const.},  
\end{equation}  
where $r$ is heliocentric distance, and $f$ is a super-radial expansion 
factor. We give radial variation of $f$ in advance. 
The structure of $B_r(r)$ is determined by surface radial field, 
$B_{r,0}$, and $f$, and is fixed during the simulations.   
In our simulations (SI05 \& SI06), we consider the magnetic field of 
$\sim$ few hundred Gauss for $B_{r,0}$ and the total super-radial expansion 
factor of 75-450, which give the average magnetic field strength of open 
field regions, $B_{r,0}/f$, of an order of 1 G. 
We should remark that recent HINODE observations found that open 
magnetic flux tubes in polar regions are connected to  ubiquitously 
distributed unipolar patches with more than kilo-Gauss \citep{tsu08b,shi10}. 
These open flux tubes super-radially open with a factor of several hundred to 
a thousand. 
Although the obtained $B_{r,0}$ and $f$ are slightly larger than those 
adopted in our simulations, the values of the average strength, $B_{r,0}/f$, 
are similar. As shown in SI06 and \citet{suz06}, $B_{r,0}/f$ is a more 
important parameter in determining the properties of the solar winds.      

We inject transverse perturbations of magnetic field with various spectra 
from the photosphere. Outgoing \Alfven waves are excited by these 
fluctuations. The propagation and dissipation of the waves are dynamically 
treated by the following MHD equations : 
\begin{equation}
\label{eq:mass}
\frac{d\rho}{dt} + \frac{\rho}{r^2 f}\frac{\partial}{\partial r}
(r^2 f v_r ) = 0 , 
\end{equation}
\begin{equation}
\label{eq:mom}
\rho \frac{d v_r}{dt} = -\frac{\partial p}{\partial r}  
- \frac{1}{8\pi r^2 f}\frac{\partial}{\partial r}  (r^2 f B_{\perp}^2)
+ \frac{\rho v_{\perp}^2}{2r^2 f}\frac{\partial }{\partial r} (r^2 f)
-\rho \frac{G M_{\odot}}{r^2}  , 
\end{equation}
\begin{equation}
\label{eq:moc1}
\rho \frac{d}{dt}(r\sqrt{f} v_{\perp}) = \frac{B_r}{4 \pi} \frac{\partial} 
{\partial r} (r \sqrt{f} B_{\perp}).
\end{equation}
$$
\rho \frac{d}{dt}\left(e + \frac{v^2}{2} + \frac{B^2}{8\pi\rho}
- \frac{G M_{\odot}}{r} \right) 
+ \frac{1}{r^2 f} 
\frac{\partial}{\partial r}\left[r^2 f \left\{ \left(p 
+ \frac{B^2}{8\pi}\right) v_r  \right. \right.
$$
\begin{equation}
\label{eq:eng}
\left. \left.
- \frac{B_r}{4\pi} (\mbf{B \cdot v})\right\}\right]
+  \frac{1}{r^2 f}\frac{\partial}{\partial r}(r^2 f F_{\rm c}) 
+ q_{\rm R} = 0,
\end{equation}
\begin{equation}
\label{eq:ct}
\frac{\partial B_{\perp}}{\partial t} = \frac{1}{r \sqrt{f}}
\frac{\partial}{\partial r} [r \sqrt{f} (v_{\perp} B_r - v_r B_{\perp})], 
\end{equation}
where $\rho$, $\mbf{v}$, $p$, $\mbf{B}$ are density, velocity, pressure, 
and magnetic field strength, respectively, and subscripts 
$r$ and $\perp$ denote radial and tangential components; 
$\frac{d}{dt}$ and $\frac{\partial}{\partial t}$ denote Lagrangian and 
Eulerian derivatives, respectively; 
$e=\frac{1}{\gamma -1}\frac{p}{\rho}$ is specific energy and we assume 
the equation of state for ideal gas with a ratio of specific heat, 
$\gamma=5/3$; 
$G$ and $M_{\odot}$ are the 
gravitational constant and the solar mass; $F_{\rm c}(=\kappa_0 T^{5/2} 
\frac{dT}{dr})$ is thermal conductive flux by Coulomb collisions, where 
$\kappa_0=10^{-6}$ in c.g.s unit \citep{brg65}; 
$q_{\rm R}$ is radiative cooling.  
We use optically thin radiative loss \citep{LM90} 
in the corona and take into account 
optically thick effects in the chromosphere \citep[][SI06]{aa89,mor04}. 

We adopt the second-order MHD-Godunov-MOCCT scheme to update the physical 
quantities. 
Each cell boundary is treated as discontinuity, and for the time evolution 
we solve nonlinear Riemann shock tube problems with the magnetic pressure term 
by using the Rankin-Hugoniot relations. Therefore, heating is automatically 
calculated from the shock jump condition. 
A great advantage of our code is that no artificial viscosity is required 
even for strong MHD shocks; numerical diffusion is 
suppressed to the minimum level for adopted numerical resolution.

We initially set static atmosphere with a temperature $T=10^4$K to see 
whether the atmosphere is heated up to coronal temperature and accelerated 
to a transonic flow. 
At $t=0$ we start the inject of the transverse fluctuations from the 
photosphere and continue the simulations until the quasi-steady states 
are achieved.  
   
\subsection{Results}

\begin{figure}
  \includegraphics[width=1.15\textwidth]{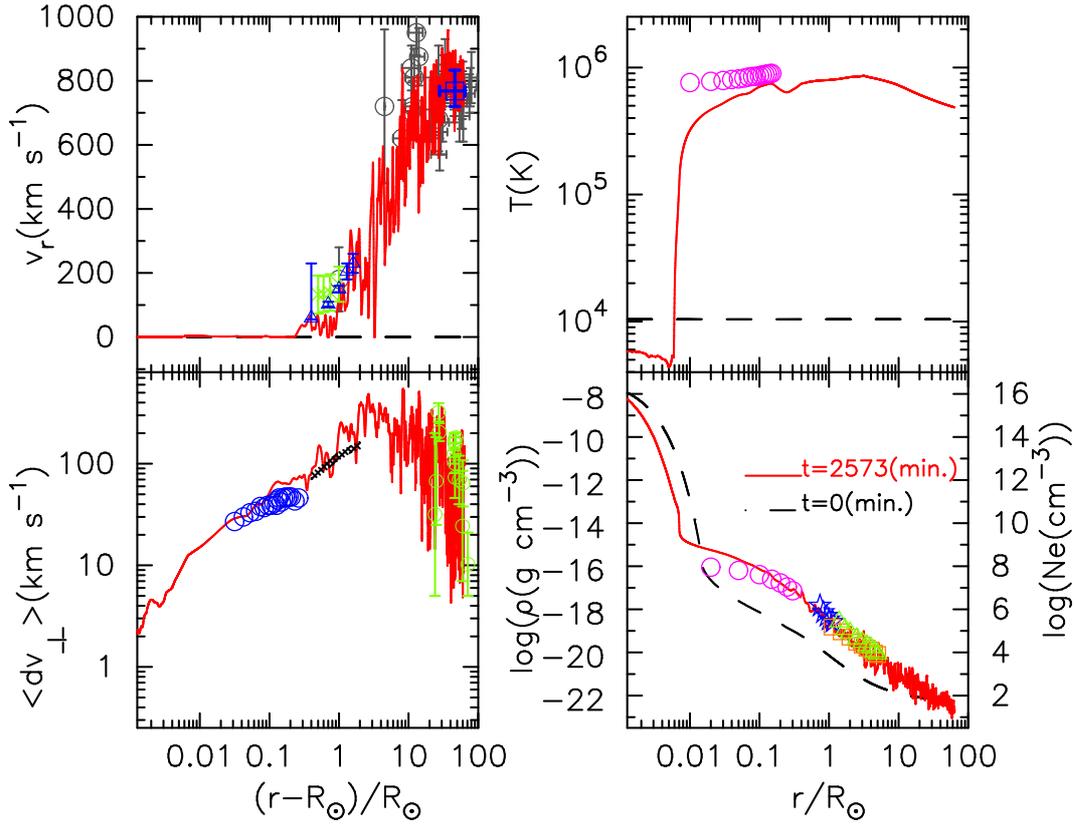}
\caption{Comparison of the simulation result with observations of fast solar 
wind. 
Outflow speed, $v_r$(km s$^{-1}$) (Top-left), temperature, $T$(K) (top-right), 
 rms transverse amplitude, $\langle dv_{\perp} \rangle$(km s$^{-1}$) 
(bottom-left), and density in logarithmic scale, 
$\log(\rho({\rm g\;cm^{-3}}))$ (bottom-right), are plotted. 
Observational data in the temperature panel are electron density, 
$\log(N_e({\rm cm^{-3}}))$ which is to be referred to the right axis. 
Dashed lines indicate the initial conditions and solid lines 
are the results at $t=2573$ minutes. In the bottom panel, the initial 
value ($\langle dv_{\perp} \rangle=0$) dose not appear. 
{\it Top-left}: Green vertical error bars are proton outflow speeds in an
interplume region by UVCS/SoHO \citep{tpr03}. 
Dark blue vertical error bars are proton outflow speeds by the Doppler 
dimming technique using UVCS/SoHO data \citep{zan02}.  
A dark blue open square with errors is velocity by IPS measurements 
averaged in 0.13 - 0.3AU of high-latitude regions \citep{koj04}. 
Light blue data are taken from \citet{gra96}; crossed bars are 
IPS measurements by EISCAT, 
crossed bars with open circles are by VLBA measurements, and 
vertical error bars with open circles are data based on observation 
by SPARTAN 201-01 \citep{hab94}. {\it top-right}: Pink circles are 
electron temperatures by CDS/SoHO \citep{fdb99}. 
{\it bottom-left}: Blue circles are non-thermal broadening inferred from 
SUMER/SoHO measurements \citep{ban98}. Cross hatched region 
is an empirical constraint of non-thermal broadening based on 
UVCS/SoHO observation \citep{ess99}. Green error bars are transverse velocity 
fluctuations derived from IPS measurements by EISCAT\citep{can02}. 
{\it bottom-right}: Circles and stars are observations by SUMER/SoHO 
\citep{wil98} and by CDS/SoHO \citep{tpr03}, respectively. 
Triangles \citep{tpr03} and squares \citep{lql97} are observations 
by LASCO/SoHO. 
}
\label{fig:fstwd}       
\end{figure}

Before discussing the reflection of the \Alfven waves in various cases,  
we explain how the coronal heating and the solar wind acceleration 
were accomplished in the typical case which we studied in SI05 and SI06 
for the fast solar wind.  
We adopt $B_{r,0} =161$ G, the total $f=75$, and the root-mean-squared (rms) 
surface amplitude, $\langle dv \rangle = 0.7$ km s$^{-1}$. 
Figure \ref{fig:fstwd} plots the final structure of the simulated solar wind 
after the quasi-steady state is achieved  
in comparison with observations of fast solar winds. 
In the four panels $v_r$(km s$^{-1}$), $T$(K), 
mass density, $\rho$(g cm$^{-3}$), and rms transverse amplitude, 
$\langle dv_{\perp} \rangle$(km s$^{-1}$) are plotted. 
As for the density, we compare our result with observed electron density, 
$N_e$, in the corona. 
When deriving $N_e$ from $\rho$ in the corona, we assume H and He are 
fully ionized, and $N_e({\rm cm^{-3}}) = 6\times 10^{23}\rho$(g cm$^{-3}$). 

Figure \ref{fig:fstwd} shows that the initially cool and static atmosphere 
is effectively heated and accelerated by the dissipation of the \Alfven waves. 
The sharp TR which divides the cool chromosphere with $T\sim 10^4$K and  
the hot corona with $T\sim 10^6$K is formed owing to a thermally unstable 
region around $T\sim 10^5$K in the radiative cooling function \citep{LM90}. 
The hot corona streams out as the transonic solar wind. 
The simulation naturally explains the observed trend quite well. 
(see SI05 and SI06 for more detailed discussions.)

\begin{figure}
  \includegraphics[width=1.\textwidth]{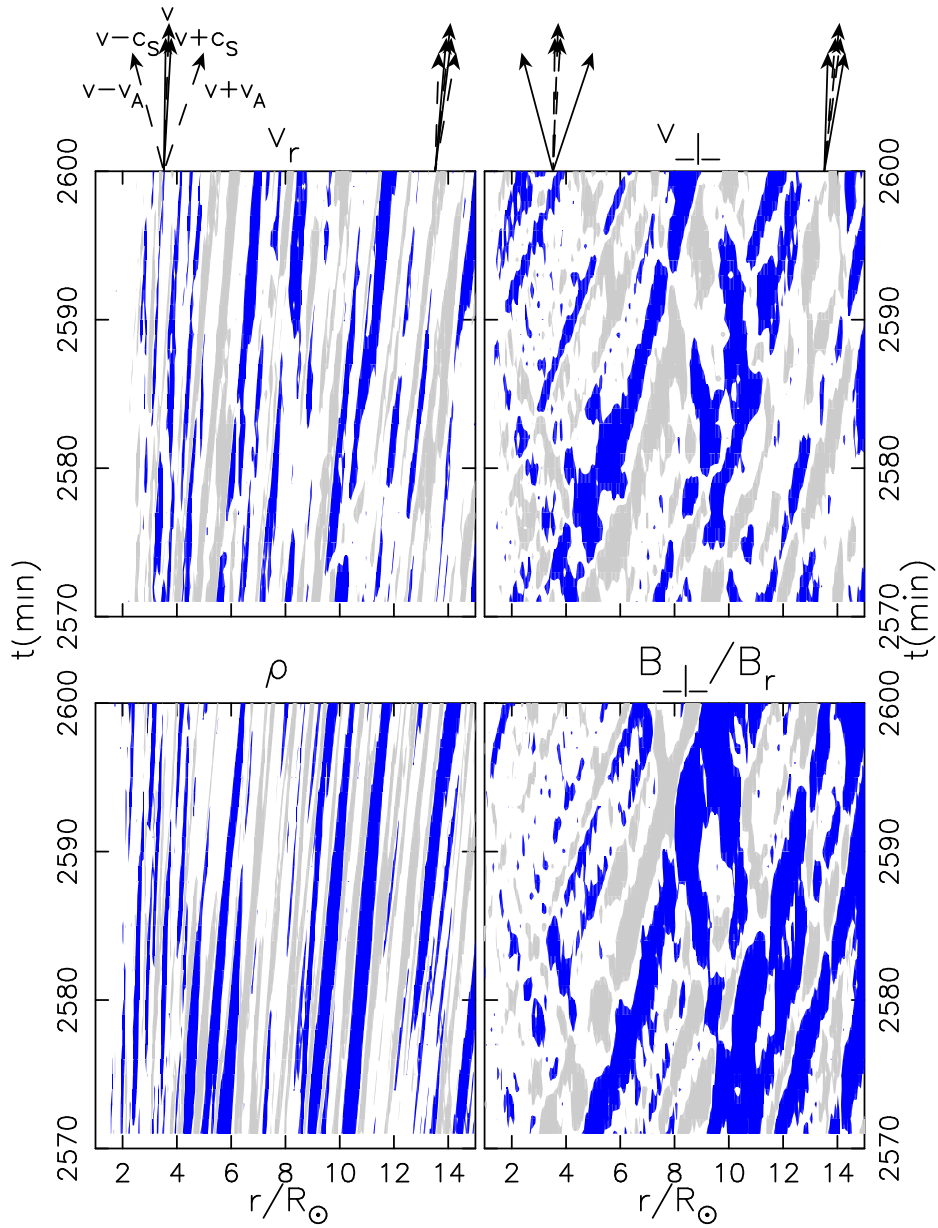}
\caption{$r-t$ diagrams for $v_r$ (upper-left), $\rho$ (lower-left), 
$v_{\perp}$ (upper-right), and $B_{\perp}/B_r$ (lower-right.) 
The horizontal axises cover from $R_{\odot}$ to $15R_{\odot}$, and 
the vertical axises cover from $t=2570$ minutes to $2600$ minutes, where 
$R_{\odot}$ is the solar radius.  
Blue and gray shaded regions indicate positive and negative amplitudes 
which exceed certain thresholds. The thresholds are $d v_r 
=\pm 96$km/s for $v_r$, $d\rho /\rho=\pm0.25$ for $\rho$, 
$v_{\perp}=\pm 180$km/s for $v_{\perp}$, and $B_{\perp}/B_r=\pm 
0.16$ for $B_{\perp}/B_r$, where $d \rho$ and $d v_r$ are 
differences from the averaged $\rho$ and $v_r$. 
Arrows on the top panels indicate characteristics of \Alfven, slow MHD 
and entropy waves at the respective locations (see text). }
\label{fig:tdd}       
\end{figure}

The reflection of \Alfven waves are easily seen in $r-t$ diagrams. 
Figure \ref{fig:tdd} 
presents contours of amplitude of $v_r$, $\rho$, $v_{\perp}$, and 
$B_{\perp}/B_r$ in $R_{\odot} \le r \le 15 R_{\odot}$ 
from $t=2570$ min. to $2600$ min. 
Blue (gray) shaded regions denote positive (negative) amplitude. Above 
the panels, we indicate the directions of the local 5 characteristics, two 
\Alfven, two slow, and one entropy waves at the respective positions. 
In our simple 1D geometry, $v_r$ and $\rho$ trace the slow modes 
which have longitudinal wave components, while $v_{\perp}$ and $B_{\perp}$ 
trace the \Alfven modes which are transverse (note that fast-mode and 
\Alfven mode degenerate in the simple 1D treatment, and then we simply 
call them \Alfven waves). 

One can clearly see the \Alfven waves in $v_{\perp}$ and $B_{\perp}/B_r$ 
diagrams, which have the same slopes with the \Alfven characteristics shown 
above. 
One can also find the incoming modes propagating from lower-right to 
upper-left as well as the outgoing modes generated from the surface\footnote{ 
It is instructive to note that the incoming \Alfven waves have the positive 
correlation between $v_{\perp}$ and $B_{\perp}$ (dark-dark or light-light 
in the figures), while the outgoing modes have the negative correlation 
(blue-gray or gray-blue).}.  
These incoming waves are generated by the reflection at the `density mirrors'  
of the slow modes in addition to the reflection owing to the shape 
deformation (SI05 \& SI06).
At intersection points of the outgoing and incoming characteristics 
the non-linear wave-wave interactions take place, which play a role 
in the wave dissipation. 

The slow modes are seen in $v_r$ and $\rho$ diagrams. Although it might 
be difficult to distinguish, the most of the patterns are due to the outgoing 
slow modes\footnote{The phase correlation of the longitudinal slow 
waves is opposite to that of the transverse \Alfven waves. 
The outgoing slow modes 
have the positive correlation between amplitudes of $v_r$ and $\rho$, 
($\delta v_r \delta \rho > 0$), while the incoming modes have the negative 
correlation ($\delta v_r \delta \rho < 0$).}
which are generated from the perturbations of the \Alfven wave 
pressure, $B_{\perp}^2/8\pi$ \citep{ks99,tsu02}. 
These slow waves steepen eventually and lead to the shock dissipation. 

The processes discussed here are the combination of the direct mode conversion 
to the compressive waves and the parametric decay instability due to 
three-wave (outgoing \Alfven, incoming \Alfven, and outgoing slow waves) 
interactions \citep{gol78,ter86} of the \Alfven waves. 
Although they are not generally efficient in the homogeneous background 
since they are the nonlinear mechanisms, the density gradient of the 
background plasma totally changes the situation. 
Owing to the gravity, the density rapidly decreases in the corona as $r$ 
increases, which results in the amplification of the wave amplitude so that 
the waves easily become nonlinear. 
Furthermore, the \Alfven speed varies a lot due to the change of the density 
even within one wavelength of \Alfven waves with periods of minutes or longer. 
This leads to both variation of the wave pressure in one wavelength  
and partial reflection through the deformation of the wave shape 
\citep{moo91}.
The dissipation is greatly enhanced by the density stratification, 
in comparison with the case of the homogeneous background. 
Thus, the low-frequency \Alfven waves are effectively dissipated,  
which results in the heating and acceleration of the coronal plasma. 
 
\begin{figure}
  \includegraphics[width=1.\textwidth]{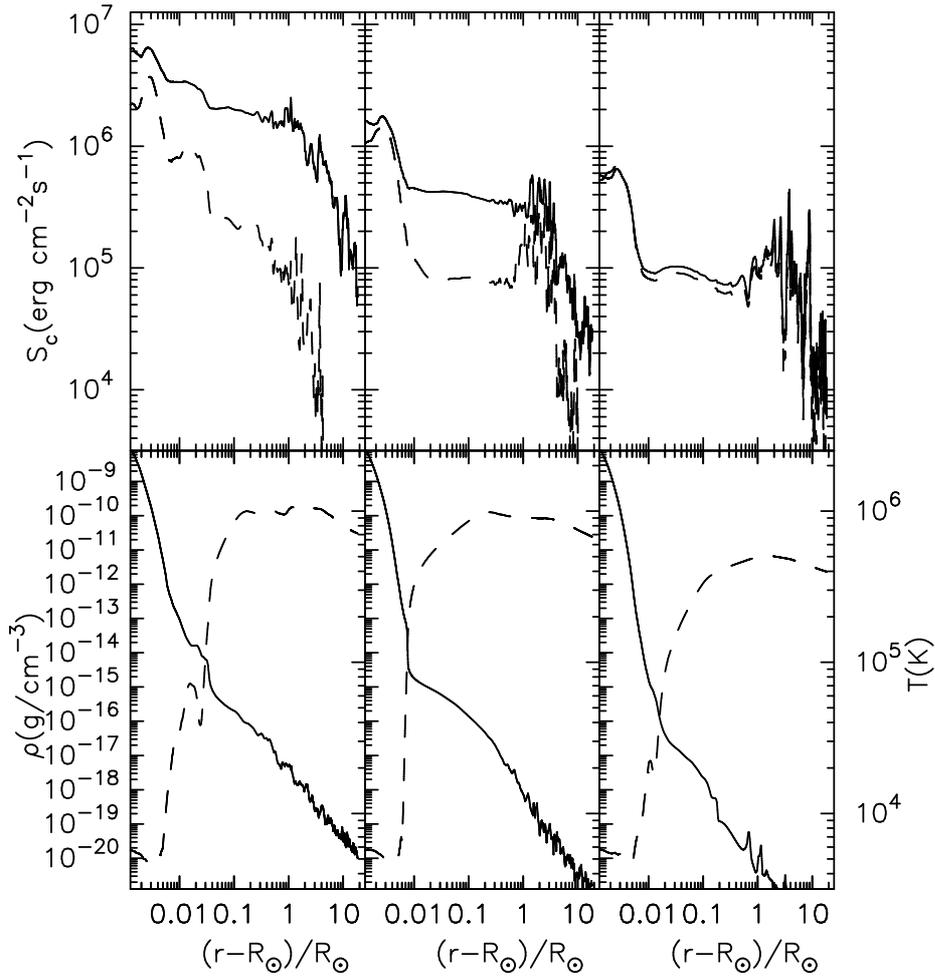}
\caption{The top panels show the wave actions ($S_{\rm c}$) of the out-going 
(solid) and incoming (dashed) \Alfven waves. The bottom panels show 
density (solid) and temperature (dashed). 
The left panels are the results of the surface 
amplitude, $\langle dv\rangle =1.4$km s$^{-1}$, the middle panels are
the results of $\langle dv\rangle =0.7$km s$^{-1}$, and the right panels 
are the results of $\langle dv\rangle =0.4$km s$^{-1}$.}
\label{fig:depdv}       
\end{figure}

From now, we examine how the reflection properties are affected by 
changing input parameters. 
In order to quantitatively study the dissipation and reflection of 
the \Alfven waves, we use an adiabatic constant, 
$S_{\rm c}$, of the outgoing \Alfven wave derived from wave action 
\citep{jaq77}: 
\begin{equation}
S_{c,\pm} = \rho \langle \delta v_{\rm A,+}^2 \rangle 
\frac{(v_r + v_{\rm A})^2}{v_{\rm A}} \frac{r^2 f(r)}{r_c^2 f(r_c)}, 
\end{equation} 
where $v_{\rm A}= B_r/\sqrt{4\pi\rho}$ is \Alfven velocity and 
\begin{equation}
\delta v_{\rm A,\pm} = \frac{1}{2}\left(v_{\perp} \mp B_{\perp}/\sqrt{4\pi\rho}
\right)
\end{equation}
is the amplitudes of the outgoing (for $+$ sign) and incoming (for $-$ sign) 
\Alfven waves (Els\"{a}sser variables). 

First, we show the results with different values of surface fluctuations, 
$\langle dv \rangle = 0.4, 0.7, 1.4$ km s$^{-1}$ (Figure \ref{fig:depdv}). 
In the upper panel, we compare $S_{c,+}$ and $S_{c,-}$ of these three cases, 
whereas we use the absolute values for $S_{c,-}$ to show in logarithmic scale. 
As shown in the bottom panels, the density is very sensitive to the input 
$\langle dv \rangle$. The density of the $\langle dv \rangle=1.4$ km s$^{-1}$ 
case is 1000 times larger than the density of the 
$\langle dv \rangle=0.4$ km s$^{-1}$ case. Accordingly the mass fluxes of the 
solar winds differs about several hundred times. 
The differences of the densities and the mass fluxes between the two 
cases are much larger than the ratio of the input energy ($\propto dv^2$) 
$\approx 12$. 
The reasons why the density sensitively depends on the input 
surface perturbation can be explained by the reflection and the nonlinear 
damping of \Alfven waves. 

The difference of the wave reflection in different $\langle dv \rangle$ 
cases are illustrated in the upper panels of Figure \ref{fig:depdv}. 
In the smaller $\langle dv \rangle$ cases, the values of the outgoing (solid) 
and incoming (dashed) components are very similar especially in the 
chromospheric regions ($r-R_{\odot}\lesssim 0.01 R_{\odot}$). 
This indicates that most of the outgoing \Alfven waves generated from the
surface is reflected back downward.
Because the heating is smaller, the temperature is lower in 
the smaller $\langle dv \rangle$ cases. Then, the scale height becomes smaller 
and the density decreases rapidly. The \Alfven speed changes 
more rapidly and the wave shape is largely deformed, which enhances 
the reflection. When the input wave energy decreases, a positive feedback 
operates; a smaller fraction of the energy can reach the coronal region. 
As a result, the density and mass flux of the solar wind becomes much 
smaller than the decreasing factor of the input energy. 

In addition to the effect of the wave reflection, the nonlinear dissipation 
of the \Alfven waves also plays a role in the sensitive behavior of 
the density on the input wave energy (SI06). When the input wave energy 
becomes smaller, the density becomes smaller as explained above. 
Then, the nonlinearity of \Alfven wave, $\delta v_{\rm A,+}/v_{\rm A}$, 
decreases not only because the amplitude, $\delta v_{\rm A,+}$, is small but 
also because the \Alfven speed, $v_{\rm A}(\propto 1/\sqrt{\rho})$, is larger. 
Therefore, the \Alfven waves do not dissipate and the heating is 
reduced, which further decreases the density; this is another type 
of positive feedbacks, which also results in the sensitive dependence 
of the density on the input wave energy.

\begin{figure}
  \includegraphics[width=1.\textwidth]{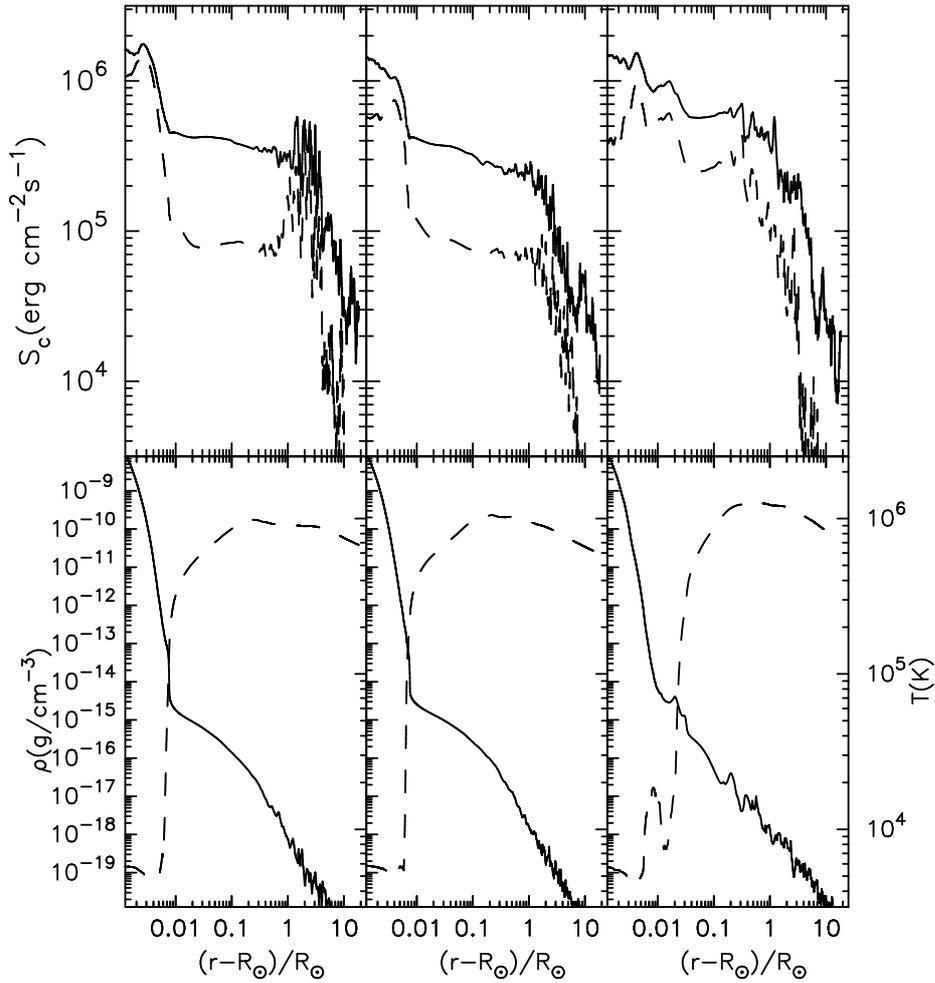}
\caption{Same as Figure \ref{fig:depdv} but for the cases with different 
spectra of surface perturbation. 
The left panels are the results of $P(\nu)\propto \nu^{-1}$ (pink noise), 
the middle panels are the results of $P(\nu)\propto \nu^0$ (white noise), 
and the right panels are the results of sinusoidal waves with 3 minutes. }
\label{fig:depsp}       
\end{figure}

Next, we investigate how the wave reflection is affected by the spectra of the 
input fluctuations. 
Here we study the three cases : The first case adopts the power spectrum, 
$P(\nu)\propto \nu^{-1}$ (pink noise), with respect to frequency, $\nu$; 
the second case adopts $P(\nu) \propto 
\nu^0$ (white noise); the third case adopts sinusoidal perturbation with 
period of 3 minutes. In the first and second cases, the range of 
the period is set from 20 seconds to 30 minutes. 
Figure \ref{fig:depsp} compares these three cases. 
The first and second cases show similar results, but in the chromospheric 
regions the first case suffers more reflection. This is because larger 
power is in longer wavelengths (lower frequency). If the wavelength is longer 
than the characteristic variation scale of \Alfven speed, \Alfven waves 
suffer more reflection. 

The third case (sinusoidal perturbation) shows considerably different 
structure from these two cases.  The transition region is not so sharp, which 
is seen in the temperature structure (dashed line in the bottom panel), 
because the heating is localized near the wave crests which are formed as
a result of steepening of nonlinearly excited compressive waves. 
Then, the heating becomes more sporadic and responding to this intermittent 
heating the transition region moves up and down. 
This is a clear contrast to the first and second cases in which the sharp 
transition regions work as 
walls against the \Alfven waves because the \Alfven speeds change drastically. 
On the other hand, in the sinusoidal case \Alfven waves become more 
transmittant owing to the fluctuating transition region. As shown in the top
panel, the reflected fraction (the ratio of the incoming component to 
the outgoing one in the chromosphere) is smaller in this case.      

\subsection{Discussions}
\subsubsection{Recent Observation of Waves}
Wave activities are observed in various portions of the solar atmosphere. 
Propagating slow MHD waves are often observed in the corona as the 
fluctuations of intensity \citep{ofm99,sak02,ban09a}. 
\citet{tom07} detected \Alfven waves propagating along with magnetic field 
lines in the corona. Although the obtained wave amplitude is smaller than 
the value required for the coronal heating, this may be due to the 
line-of-sight effect \citep{mci08}.  Indeed, nonthermal broadening of spectral 
lines, which are supposed to be closely linked to \Alfven waves, gives 
much larger amplitudes \citep{sin04,ban09b}.   

\Alfven(ic) waves are detected at the photospheric level in quiet regions 
\citep{ulr96} by analyzing the phase correlation of magnetic field and 
velocity perturbations. Similar attempts have been carried out in sunspot 
regions as well \citep{lit98,bel00},   
Recently, \citet{ft09} carried out very detailed analysis of the observed 
fluctuations  
by the spectro-polarimeter (SP) of the Solar Optical Telescope 
\citep[SOT;][]{tsu08,sue08,ich08,shi08} aboard the HINODE satellite. 
They obtained the fluctuations of magnetic field, $\delta B$, the 
fluctuations of velocity field, $\delta v$, and the fluctuations of 
intensity, $\delta I$, 
which reflects density perturbation, $\delta \rho$ of pores, and magnetic 
concentrations in plages. The periods of the oscillations are
distributed from 3 to 6 minutes in the pores and from 4 to 9 minutes in 
the plages. 
From the phase correlation of $\delta B$, $\delta v$, and $\delta I$ 
they investigated the modes and directions of the waves to that are attributed
the observed fluctuations.  
One of the most important results is that the phase differences between 
$\delta v$ and $\delta B$ are nearly -90 degrees \footnote{\citet{ft09} 
picked up the regions with positive magnetic polarity (magnetic fields toward 
observers), which is the reason why there are no data with the phase shift of 
$+90$ degrees. Data with negative polarity actually show the phase shift peaked 
around $+90$ degrees (Fujimura \& Tsuneta, private 
communication). }. The obtained phase correlation is consistent with 
standing kink-mode (transverse) or sausage-mode (longitudinal) waves, 
which are, roughly speaking, surface-mode counterparts of \Alfven and 
slow MHD waves.  
Namely, waves that propagate to positive and negative directions are almost 
equally exist at the photospheric level.    
Although it is quite difficult to estimate the fraction of sausage 
(compressive) mode and kink (nearly incompressive) mode because the 
transformation from $\delta I$ to $\delta \rho$ is not straightforward, 
it is expected that regions with smaller $\delta I$ are dominated by 
kink-mode oscillations. Based on this consideration, they picked up a 
particular pore region with relatively small $\delta I$. 
In this region, the phase difference between $\delta B$ and 
$\delta v$ is -96 degrees, which indicates that the upward propagating flux is 
slightly larger than the downward propagating component. Assuming the observed 
fluctuations attribute to kink-mode waves, they obtained the net Poynting flux 
that propagates upward as $2.7\times 10^6$ erg cm$^{-2}$s$^{-1}$. 

Interestingly enough, our simulations also show very effective reflection 
of \Alfven waves below the transition region. 
In the reference case for the fast solar wind (Figure \ref{fig:fstwd}),  
85\% of the initial upward Poynting flux of the \Alfven waves from the 
photosphere is reflected back downward before reaching the corona. 
This implies that one can observe the comparable amount of downward flux 
to the upward component.
The net leakage of the upgoing flux to the corona in this simulation is 
15\% of the input at the photosphere ($\sim 5\times 10^5$ erg cm$^{-2}$s$^{-1}$).
 These features seem quite consistent, 
at least in a qualitative sense, with the observation by \citet{ft09}. 
The specific values of the leaking and reflected fluxes
depend on strength and configuration of magnetic flux tubes and amplitudes 
of surface fluctuations.  

\subsubsection{Data Driven Simulations}

The observation data are also used in numerical simulations. Recently, 
various groups are performing so-called data-driven simulations 
by adopting  magnetic field data on the Sun \citep[e.g.][]{man08,kat09}. 
While these works are mainly aiming at global heliospheric phenomena such as 
propagation of coronal mass ejections,  
observed data by HINODE are also applied for MHD simulations of surface 
activities.  
\citet{ms09} have carried out numerical simulations in open field regions 
by using observed transverse motions on the Sun. 
They obtained horizontal motions of granules by local correlation tracking
\footnote{To do so HINODE/SOT is the best telescope, because horizontal 
oscillations are largely affected by the atmospheric seeing.}. 
These transverse motions are expected to excite \Alfvenic waves that 
propagate along the vertical magnetic fields. 
They observed 14 different regions and derive the power spectrum of the 
transverse oscillations. Using the obtained spectrum as the input from the 
photosphere, they performed 1D MHD simulations in open field regions up to 
the corona following \citet{ks99}. 
They found that, compared to artificial inputs of white noise or pink noise, 
more fraction of the \Alfven wave energy transmits 
through the transition region. They interpret that this is because more energy 
is resonantly trapped between the photosphere and the transition region and 
eventually leaks into the corona.

\subsubsection{Slow Solar Winds}
In this paper, we have introduced our simulation results for 
the fast solar winds from polar coronal holes. 
In SI06 and \citet{suz06}, we tried to explain observed data of 
slow solar winds in the same framework of the nonlinear dissipation of 
the \Alfven waves. We have shown that the slow winds from flux tubes 
with smaller $B_{r,0}/f$ and slightly larger $\langle dv_{\perp} \rangle$ 
well explain the observed properties of the slow streams from mid- to 
low-latitude regions; the same physical mechanism operates but the 
environments (e.g. geometries of flux tubes) make the differences between 
the fast and slow solar winds. 
This interpretation is quite consistent with the observed data by 
Interplanetary scintillation measurements \citep{kfh05}. 

HINODE observations detected mass outflows from open regions in the 
vicinities of active regions \citep{sak07,ima07,har08}. These outflows 
might become slow solar winds and possibly give the significant 
contribution to the total mass loss from the Sun \citep{sak07}. 
What is puzzling is that the acceleration seems to take place at a very low 
altitude. 
\citet{harr08} measured the Doppler velocities of the outflows and 
concluded that the outflow speeds reach $\approx 100$ km s$^{-1}$ at 
the height of $\sim 0.1 R_{\odot}$ from the surface. 
Such rapid acceleration is very different from those inferred in the 
classical slow winds. 
For instance, the observations by \citet{she97} shows that 
the slow winds associated with the streamer belt reach $\simeq 100$ km s$^{-1}$ 
above 2-3 solar radii. 

It is difficult to explain the observed rapid acceleration by 
our simulations of the nonlinear \Alfven waves. If we increase the input 
energy ($\sim \langle dv^2\rangle$), density, rather than outflow 
speeds, increases owing to the larger heating. Probably other calculations 
or simulations will give similar tendency. 
To explain the outflow from the near active regions, pure momentum inputs 
are necessary.

\section{Evolution to Red Giants}
\label{sec:stw}
When the Sun evolve to a red giant star, the properties of the stellar 
wind also change. 
When the stellar atmosphere becomes cool enough to form dusts during 
the asymptotic giant branch (AGB), the stellar wind is mainly driven by the 
radiation pressure on the dust particles \citep{bow88}. 
Before the AGB phase, namely from the main sequence to 
(early) red giant branch (RGB), the driving mechanism of the stellar winds 
is regarded to be similar to the solar wind; the origin of the driver is the 
kinetic energy of the surface convection, and MHD waves generated by the 
turbulent motion of the surface convection accelerates the stellar winds 
\citep[e.g.][]{hm80,jo89,cm95}. 
However, the properties of the stellar 
winds change with stellar evolution primarily because the surface gravity 
decreases. 

\begin{figure*}
  \includegraphics[width=0.8\textwidth]{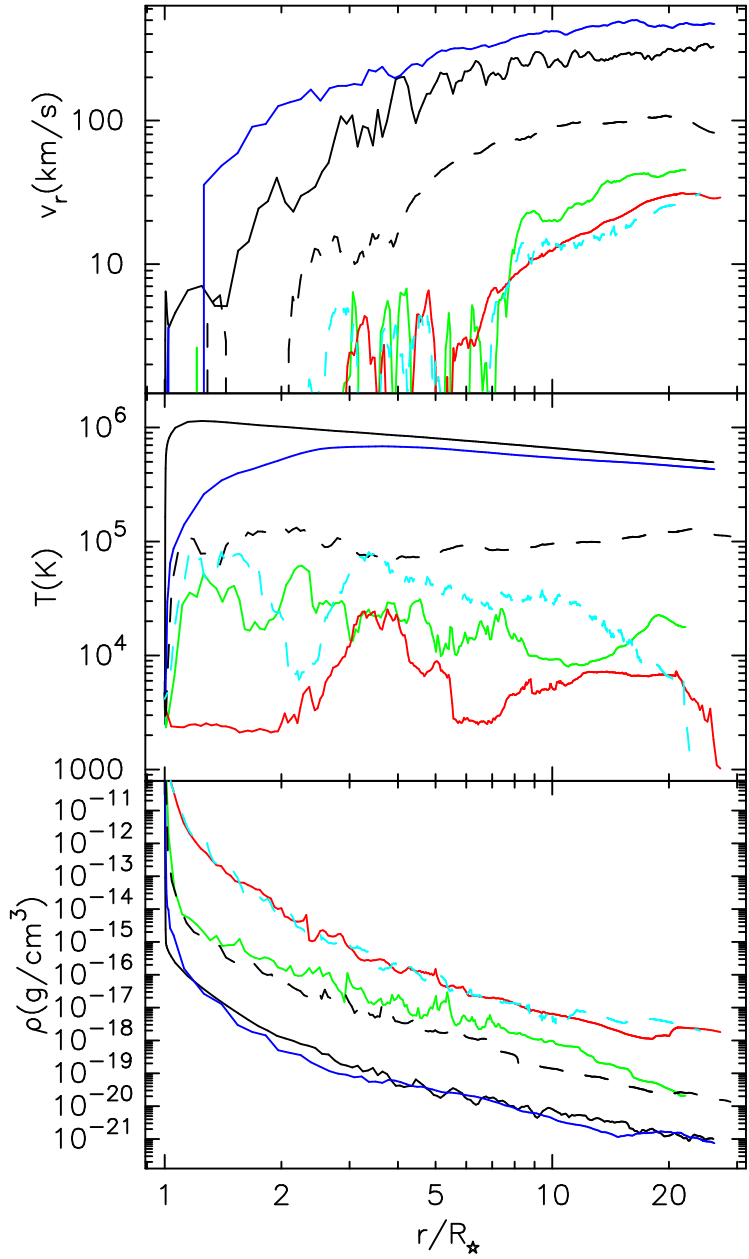}
\caption{Evolution of stellar wind from the main sequence to RGB. 
From top to bottom, radial outflow velocity, $v_r$ (km s$^{-1}$),  
temperature, $T$ (K), and density, $\rho$(g cm$^{-3}$), are plotted. 
The solid lines are the results 
of the 1 $M_{\odot}$ stars ; the black, blue, green, and red lines are the 
results of the star with the surface gravity, $\log g=4.4$ (Sun), 3.4, 2.4, 
and 1.4, respectively.  
The dashed lines are the results of the $3\; M_{\odot}$ stars; the black and 
light-blue lines are the results of the stars with the surface gravity, 
$\log g=2.4$ and 1.4.  }
\label{fig:evl}       
\end{figure*}

Based on these considerations, we studied the evolution of the stellar 
winds with the stellar evolution from the main sequence to the RGB (S07). 
We extend the solar wind simulation introduced in 
the previous section to simulate the red giant winds. 
We change the stellar radius, which controls the surface gravity, and 
effective temperature (or sound speed). 
We input the surface fluctuations that excite outgoing waves from the 
photosphere in the same manner as in the solar wind simulation. 
The amplitudes and spectra of the perturbations are estimated 
from the scaling relation of surface convective flux and period 
\citep[][see also Brun \& Palacios 2009 for recent simulations]{ren77}. 
We simulated the stellar winds from a $1M_{\odot}$ star with different 
surface gravities, $\log g=4.4$, 3.4, 2.4, and 1.4 (the Sun has $\log g=4.4$) 
to investigate the effect of the stellar evolution. In addition, we simulated 
the stellar winds from a $3M_{\odot}$ star with $\log=2.4$ and 1.4.  
In all the models we use the photospheric magnetic field, $B_{r,0}=240$ G, and 
the total super-radial expansion factor, $f=240$. Note that as for 
the main sequence star (the Sun in this case) these values are between the 
models for fast and slow solar winds (SI06). 
The stellar wind structures after the quasi-steady states are presented 
in Figure \ref{fig:evl}.

\begin{figure}
  \includegraphics[width=0.8\textwidth]{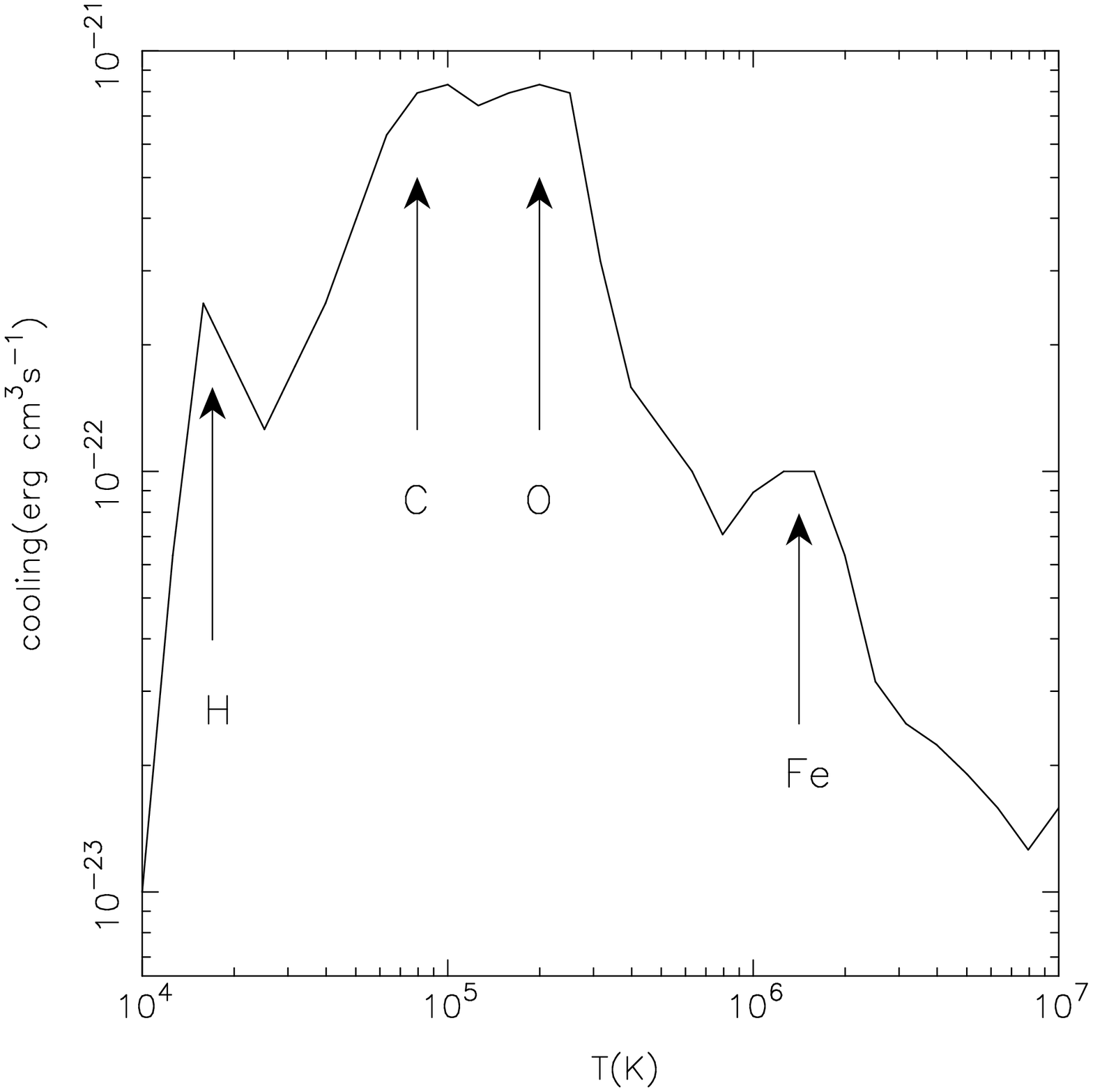}
\caption{Cooling function of the solar metallicity gas from \citet{LM90}. 
The main coolants in the respective temperature ranges are also indicated with 
arrows.}
\label{fig:clg}       
\end{figure}

The figure shows that hot coronae with $T\gtrsim 10^6$ disappears when 
the star evolves to $\log g \le 2.4$, and the density in the atmospheres and 
winds increase (when measured in stellar radii). 
The increase of the density is owing to the decrease of the surface 
gravity as a result of the expansion of the star; more mass can be lifted 
up to upper regions because of the smaller gravity. 
The decrease of the temperature can also be partly explained by the decrease 
of the gravity. 
When the star evolves to 
the red giant, the escape velocity becomes comparable to the sound speed; 
for example, the sound speed of the coronal gas ($\sim 150$ km s$^{-1}$) 
exceeds the escape speed of the red giant stars with $\log g \sim 2$ at a few 
stellar radii. Then, the hot corona with $T\gtrsim 10^6$ K cannot be confined  
by the gravity and inevitably streams out.

While the increase of the density is rather continuous with the stellar 
evolution, the temperature rapidly drops from the star with $\log g=3.4$ 
to the star with 2.4. 
Thermal instability plays a role in this rapid decrease of the temperature, 
in addition to the gravity effect explained above. 
Figure \ref{fig:clg} shows the radiative cooling function adopted from 
\citet{LM90} of the solar metallicity gas under the optically thin 
approximation, which we use in the simulation. 
As shown in the figure, the radiative flux decreases for increasing 
temperature in $10^5\;({\rm K}) \lesssim T \lesssim 10^6\;({\rm K})$. 
In this region, the cooling of gas is suppressed when it is heated up; 
it is thermally unstable. Then, the gas is stable in only $T\lesssim 10^5$ K 
or $T\gtrsim 10^6$ K in which thermal conduction also plays a role in 
the stabilization. 
This is the main reason why the temperature drops 
from $\sim 10^6$ K to $\lesssim 10^5$ with stellar evolution.
For more detail please see S07.

\begin{figure}
  \includegraphics[width=1.\textwidth]{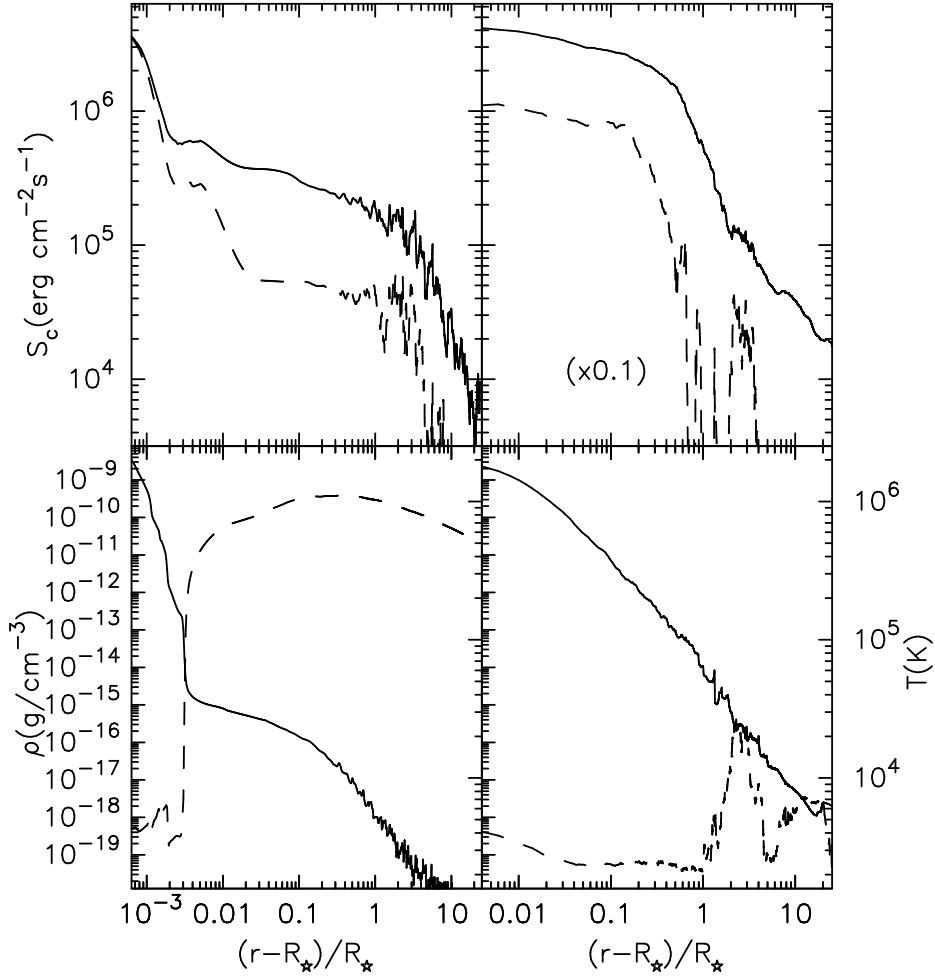}
\caption{The left panels are the results 
of the main sequence star with the surface gravity of $\log g=4.4$ and 
the right panels are the results of the red giant star with $\log g =1.4$ 
of 1 $M_{\odot}$ star. 
The top panels show the wave actions ($S_{\rm c}$) of the out-going 
(solid) and incoming (dashed) \Alfven waves. The bottom panels show 
density (solid) and temperature (dashed). The wave action ($S_{\rm c}$) 
of the red giant case is reduced to 0.1 times of the original value for 
comparison. }
\label{fig:depev}       
\end{figure}

The properties of the reflection of the \Alfven waves are also 
affected by the stellar evolution. 
Figure \ref{fig:depev} compares the reflection of the outgoing \Alfven waves 
between the main sequence star ($\log g=4.4$) and the moderately evolved 
red giant star ($\log g=1.4$). 
As clearly illustrated in the bottom panels, the reflection of the outgoing 
\Alfven waves is significantly suppressed in the evolved star compared 
to the main sequence star. 
only 30\% of the input \Alfven waves from the surface is reflected back in 
the evolved $1 M_{\odot}$ star with $\log g=1.4$, 90\% is reflected back 
in the main sequence star.   
This is mainly because the density slowly decreases 
in the red giant star and the variation scale of the \Alfven speed is larger. 
Then, the outgoing \Alfven waves do not suffer reflection so much and 
more fraction of the input energy can reach higher altitudes. 
However, because of the larger density the radiative loss is more effective 
in the evolved stars (S07). 

The mass loss rate, $4\pi \rho v_r r^2$, of the red giant star with 
$\log g=1,4$ of 1 $M_{\odot}$ star is more than $10^5$ times larger 
than the mass loss rate of the main sequence star. This is much larger 
than the increase of the stellar surface area ($=1000$ times); the increase 
of the mass flux, $\rho v_r$, itself contributes significantly 
\citep[S07; see also][]{sc05}. 


\begin{figure}
  \includegraphics[width=1.\textwidth]{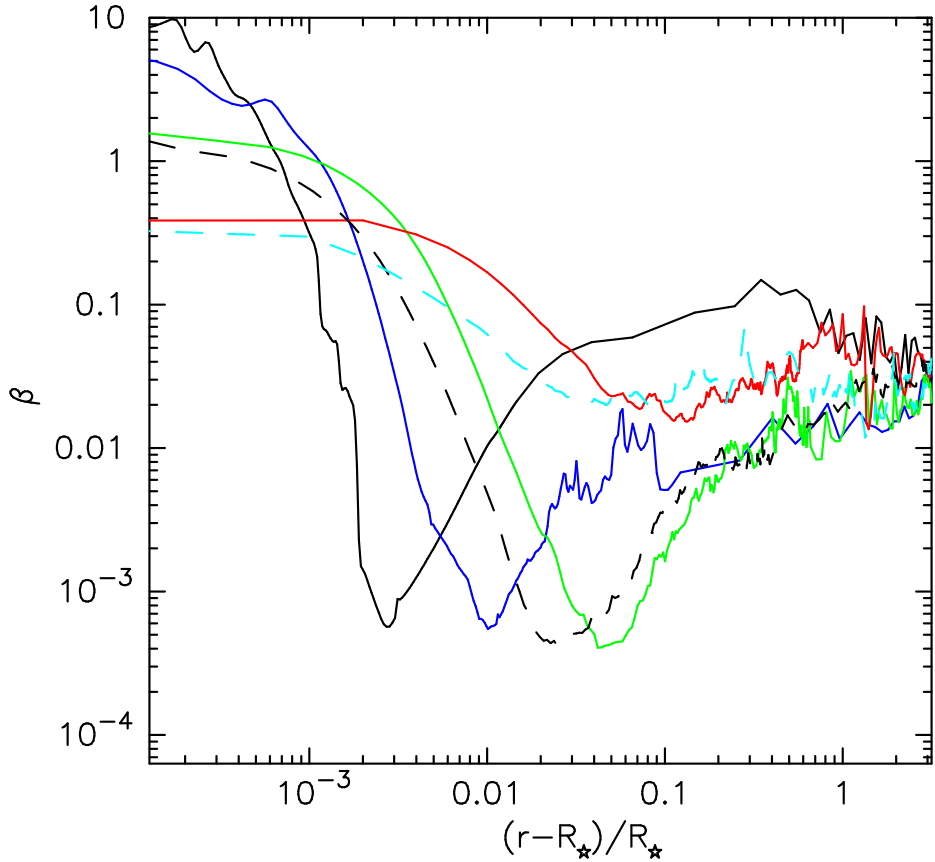}
\caption{Plasma $\beta$ values as functions of stellar radii. 
The solid lines are the results of the $1 M_{\odot}$ star; the black, blue, 
green, and red lines are the 
results of the star with the surface gravity, $\log g=4.4$ (Sun), 3.4, 2.4, 
and 1.4, respectively.  
The dashed lines are the results of the $3\; M_{\odot}$ stars; the black and 
light-blue lines are the results of the stars with the surface gravity, 
$\log g=2.4$ and 1.4.   }
\label{fig:beta}       
\end{figure}

The \Alfven wave-driven stellar wind plays an important role in terms 
of the energy conversion from magnetic energy to other types of energy. 
To study the energy conversion, a plasma $\beta$ value, which is 
defined as the ratio of gas pressure to magnetic pressure, 
\begin{equation}
\beta=8\pi p / B^2, 
\end{equation}
is a useful parameter. 
Figure \ref{fig:beta} shows the $\beta$ values for the different six 
models\footnote{
This figure is modified from the bottom panel of Figure 11 of S07. 
In S07, the $\beta$ values were estimated from the radial magnetic 
field strength, $B_r$. In this paper, we calculate $\beta$ from the total 
magnetic field strength, the sum of radial and transverse components.}. 
In the region closed to the surface, the $\beta$'s decrease on $r$ at first, 
which is more clearly seen in the unevolved stars. 
In this region, the atmosphere is mostly static and the density decreases 
rapidly according to an exponential manner. The decrease of the density is 
faster than the decrease of the magnetic energy $\propto B^2$, and then, 
the atmosphere becomes magnetically dominated, $\beta <1$. 
If the static atmosphere continued to the upper altitudes without the 
dissipation of the \Alfven waves, the $\beta$ values would keep decreasing.     

In reality, however, the density structure is redistributed by the heating 
from the wave dissipation. In the main sequence ($\log g=4.4$) and subgiant 
($\log g=3.4$) stars, 
the hot coronae are formed, which give larger pressure scale heights (
$\propto$ temperature). Therefore, the densities decrease more slowly 
than the magnetic energy, and the $\beta$ values increase in the coronal 
regions. 
In the red giant stars, although hot coronae do not form as in the main 
sequence and subgiant stars, the gas is lifted up directly by magnetic 
pressure associated with \Alfven waves in the weak gravity conditions, 
which are gradually connected to the stellar wind regions. 
Interestingly, the final $\beta$ values 
are $0.01-0.1$, being independent from the surface gravities. 
The redistribution of the density through the wave dissipation makes the 
plasma $\beta$ values stay at the moderate values.


\section{Turbulent-driven Accretion Disk Winds}
\label{sec:dsw}
\begin{figure}
  \includegraphics[width=1.\textwidth]{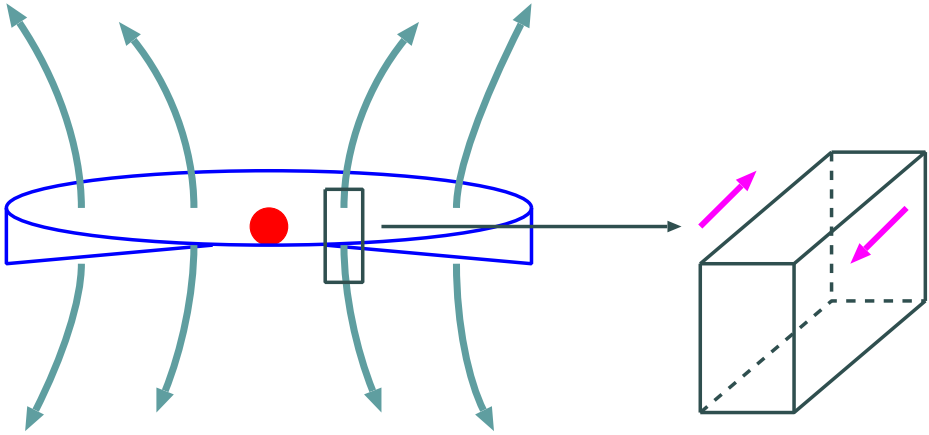}
\caption{Schematic picture of the local shearing box simulation of an 
accretion disk.}
\label{fig:schdsk}       
\end{figure}

We investigated accretion disk winds that are driven by MHD turbulence 
in accretion disks \citep{si09}. 
The geometrical aspects of disk winds are different from those of the stellar 
winds as easily expected. 
On the other hand, the driving mechanism shows similarities to the solar 
and red giant winds, which we discuss from now.  

To study turbulence driven accretion disk winds, we performed 3D MHD 
simulations in a local patch of an accretion disk (Hawley et al. 1995; 
Matsumoto \& Tajima 1995; Figure \ref{fig:schdsk}). 
Here, we neglect the effects of curvature and the simulations are carried 
out in a local Cartesian box that co-rotates with a Keplerian rotating 
disk. Following the convention (Hawley et al. 1995), $x$, $y$, and 
$z$ coordinates corresponds to radial, azimuthal, and vertical directions, 
respectively. While in the $x$ direction we do not take into 
account the radial gradients of the background density, in the $z$ direction 
the density stratification owing to the gravity by a central star is included. 
The size of the simulation box is $(x,y,z)=(\pm 0.5 H, \pm 2H, \pm 4H)$, where
a scale height, $H$, is defined from rotation frequency, $\Omega$, 
and sound speed as $H=\sqrt{2}c_s/\Omega$. 
We use $(32,64,256)$ mesh points for the $(x,y,z)$ coordinates.
The $z$ size of the simulation box covers from the sufficiently lower 
region to upper region of an entire disk so that we can simultaneously 
treat the amplification of magnetic field by MRI and excitation of disk 
winds in a self-consistent manner.      
This is an advantage of the simulation of the accretion disk winds in 
comparison to the simulations of solar/stellar winds, in which we give the 
conditions at the surface based observationally inferred values without 
considering the generation of the magnetic field in the solar/stellar 
interiors. 

Because the co-rotating box is not in an inertial coordinate, inertial forces 
need to be taken into account in the momentum equations :   
\begin{equation}
\frac{d\mbf{v}}{d t} = -\frac{1}{\rho}\nabla(p+\frac{B^2}{8\pi}) 
+ \frac{(\mbf{B}\cdot\nabla)\mbf{B}}{4\pi\rho} 
- 2\mbf{\Omega_0}\times\mbf{v} + 3\Omega_0^2 \mbf{x} 
- \Omega_0^2 \mbf{z},
\label{eq:mom}
\end{equation}
where the third term on the right hand side is Coriolis force, the fourth 
term corresponds to tidal expansion which is the sum of centrifugal 
force and the radial direction of the gravity by a central star, and the 
last term is the vertical component of the gravity which gives the 
stratification of the density in the vertical direction. 
We assume isothermal gas instead of solving the energy equation in order 
to focus of the dynamics of the disk winds. 
In the $x$ direction, the shearing boundary condition (Hawley et al.1995) 
is adopted to mimic the differential rotation of a Keplerian rotating disk. 
The simple periodic boundary condition is applied to the $y$ direction. 
The outgoing condition (SI06) is prescribed to the $z$ direction to 
properly treat streaming out disk winds.  
We would like to note that our simulation is the first attempt that adopts 
the real outgoing boundary to the disk wind simulation in the local shearing 
box, whereas the simple zero-gradient boundary was used in previous 
studies \citep{ms00,hir06}.  
As the initial condition, we give Keplerian rotation, $v_x =-3/2\Omega 
x$, with small perturbations as seeds for MRI and set up weak vertical 
magnetic fields with $\beta=10^6$ at the midplane. In the vertical direction, 
we give the hydrostatic density structure, $\rho \propto \exp(-z^2/H^2)$. 

\begin{figure}
  \includegraphics[width=1.\textwidth]{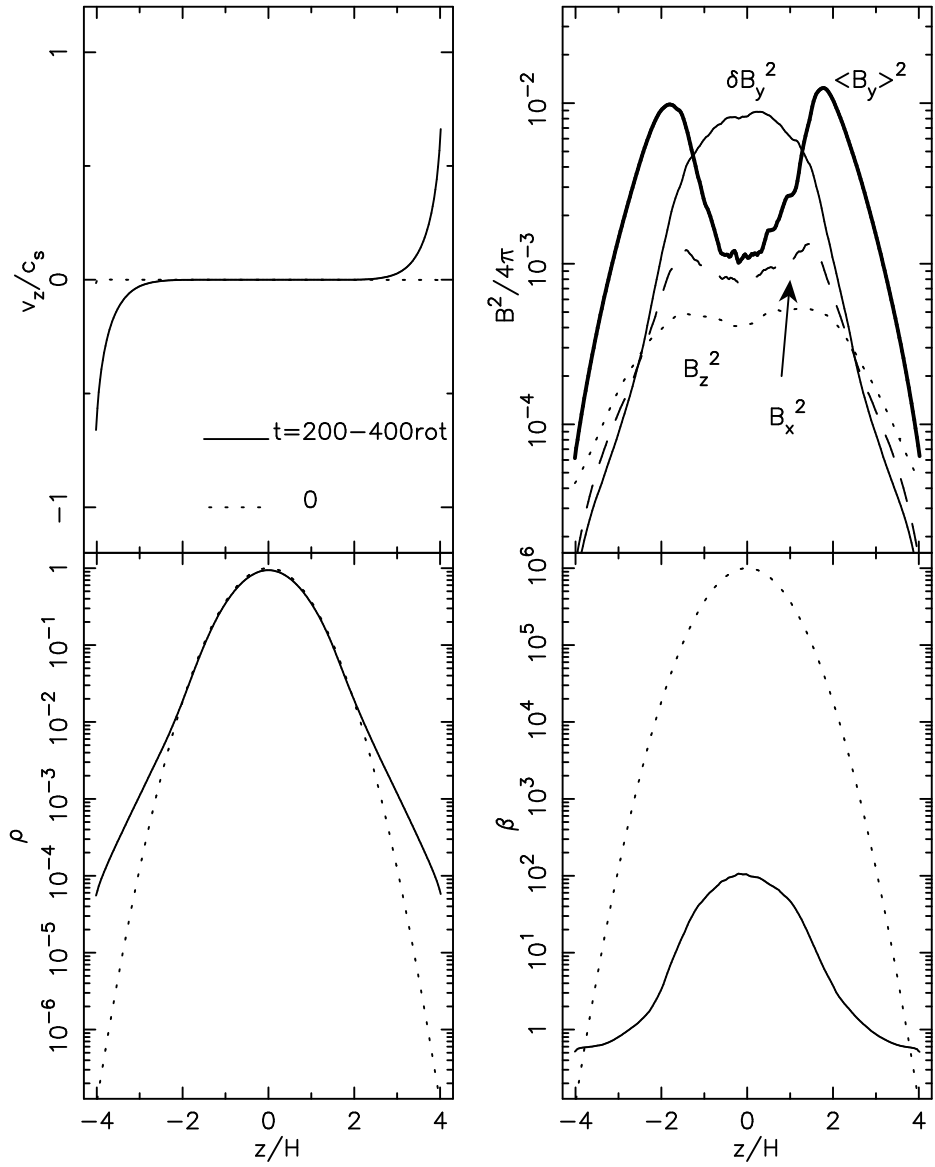}
\caption{Time-averaged disk structure during $t=200-400$ rotations. 
The variables are also averaged on $x-y$ plane at each $z$ grid. 
The top left panel shows $v_z/c_s$ (solid), whereas the dotted line is the 
initial condition ($v_z/c_s=0$). 
The bottom left panel presents the density (solid) in comparison with the 
initial condition (dotted). The top right panel 
presents magnetic energy, $B^2/4\pi$. The dashed, solid, and dotted lines 
correspond to $x$, $y$, and $z$ components, whereas the $y$ component 
shows both mean (thick) and fluctuation (thin) components. 
The bottom right panel shows the plasma $\beta$ (solid) in comparison with 
the initial value (dotted). 
}
\label{fig:zav}       
\end{figure}

In this review paper we mainly focus on the time-averaged 
structure of disk winds. We would like to briefly describe the time evolution 
and dynamical properties of the simulated local accretion disk 
(see Suzuki \& Inutsuka 2009 for more detail).  
After $\sim 3$ rotations, MRI triggers MHD turbulence firstly around 
$z\sim \pm (2-3) H$. In the higher regions ($|z|>3H$), MRI does not set in 
because the circumstance is magnetically dominate from the beginning and 
stable against MRI. On the other hand, the delay of the MRI trigerring 
in the near-midplane region is an artifact. Since in this region 
the plasma $\beta$ is too high (gas pressure dominated) at first, we cannot 
resolve the most unstable wavelength of MRI. However, MRI gradually takes 
place later on. As the magnetic field strength increases, the most unstable 
wavelength can be eventually resolved at the midplane. 
After 100 rotation, the entire region except the surface 
regions ($|z|>3H$) becomes turbulent and the disk winds are driven by 
MHD turbulent pressure from the upper and lower surfaces. At $200$ rotations, 
the magnetic field strength almost saturates, balancing the amplification 
by MRI with the cancellation by magnetic reconnections and the escape with 
the disk winds\footnote{Since the ideal MHD condition is assumed in the 
simulation, the magnetic reconnections take place due to the numerical 
resistivity determined by the grid scale. 
Then, the saturation level might depend on the resolutions of simulations, 
whereas this problem is still under debate \citep{san04,pcp07}.}. At this time, 
the magnetic energy is amplified 1000 times of the initial energy of the weak 
vertical field, and the field lines are dominated by the toroidal ($y$) 
component (see below).

An interesting feature is that the disk winds are blow off intermittently. 
The mass fluxes of the disk winds become strong every 5-10 rotations 
quasi-periodically. This is a consequence of breakups of large scale channel 
flows at $z\approx \pm 2H$ (Suzuki \& Inutsuka 2009).  

Figure \ref{fig:zav} presents the disk wind structure averaging over 200 - 400 
rotations. The variables are averaged on the $x-y$ plane at each $z$ point. 
The top left panel shows that the gas streams out of the upper and lower 
surfaces. The average outflow velocity is nearly the sound speed at the upper 
and lower boundaries. The disk winds are dominantly accelerated by magnetic 
energy (Poyinting flux) of the MHD turbulence, which we inspect from now. 

The top right panel shows magnetic energy at the saturated state. 
The dashed, solid, and dotted lines are $x$, $y$, and $z$ components. 
In the $y$ component we are showing both mean, $\langle B_y\rangle^2$ and 
fluctuation, $\delta B_y^2$, components. $\langle B_y\rangle^2$ 
is the simple average on $x-y$ planes, $\langle B_y(z)\rangle=\int\int 
dx dy B_y(x,y,z)/(L_{x}L_{y})$, and the the fluctuations are 
determined from $\delta B_y^2(z)=\int\int dx dy(B_y(x,y,z)-\langle 
B_y(z)\rangle)^2/(L_{x}L_{y})$, where $L_x(=H)$ and 
$L_y(=4H)$ are the $x$ and $y$ lengths of the simulation box. 
As for $B_x$ and $B_z$ the fluctuation components greatly dominate the means, 
and so we simply present $B_x^2$ and $B_z^2$.  
The magnetic energy, which is dominated by the toroidal ($y$) component 
as a consequence of winding, is amplified by $\approx$ 1000 times of
the initial value ($B_{z,0}^2/4\pi=10^{-6}$) in most of the region 
($|z|<3H$). 
While in the region near the mid-plane ($|z|<1.5H$), 
the magnetic field is dominated by fluctuating component ($\delta B_y$), 
the coherent component ($\langle B_y \rangle$) dominates in the regions 
near the surfaces ($|z|>1.5H$). 
In the surface regions the magnetic pressure 
is comparable to or larger than the gas pressure ($\beta\lesssim 1$), 
and thus, the gas motions cannot control the configuration of 
the magnetic fields. 
Therefore, the field lines tend to be straightened by magnetic tension 
to give $\langle B\rangle^2>\delta B^2$ there, 
even if the gas is turbulent.  
We also note that $\langle B_z^2\rangle $ 
 is amplified by MRI and Parker (1966) instability, 
 whereas $\langle B_z\rangle^2$ is strictly conserved.

The comparison of the final density structure (solid) with the initial 
hydrostatic structure (dotted) in the bottom left panel shows that the 
mass is loaded up to the onset regions of outflows from $z\approx\pm 2H$ 
by the amplified magnetic energy. 
The plasma $\beta$ value in the bottom right panel is a good indicator to 
understand the locations of the mass loading.  
The panel shows that the wind onset regions correspond to $\beta\approx 1$, 
which indicates that the winds start to be accelerated when the magnetic 
pressure exceeds the gas pressure. 
In the further upper regions, $|z|\gtrsim 3H$, $\beta$ stays almost constant 
slightly below unity. Without disk winds, $\beta$ decreases with increasing 
height in the hydrostatic structure. However, the mass loading and disk winds 
inhibit the decrease of $\beta$.  
We can interpret that the density structure is redistributed to give the almost 
constant $\beta$ in the wind regions. This is, in a qualitative sense, 
similar to the $\beta$ structures in the stellar winds (Figure \ref{fig:beta}). 
The detailed features are different mainly because we do not solve the energy 
equation with radiative cooling and conduction in the disk wind simulations.

\begin{figure*}
  \includegraphics[width=0.8\textwidth]{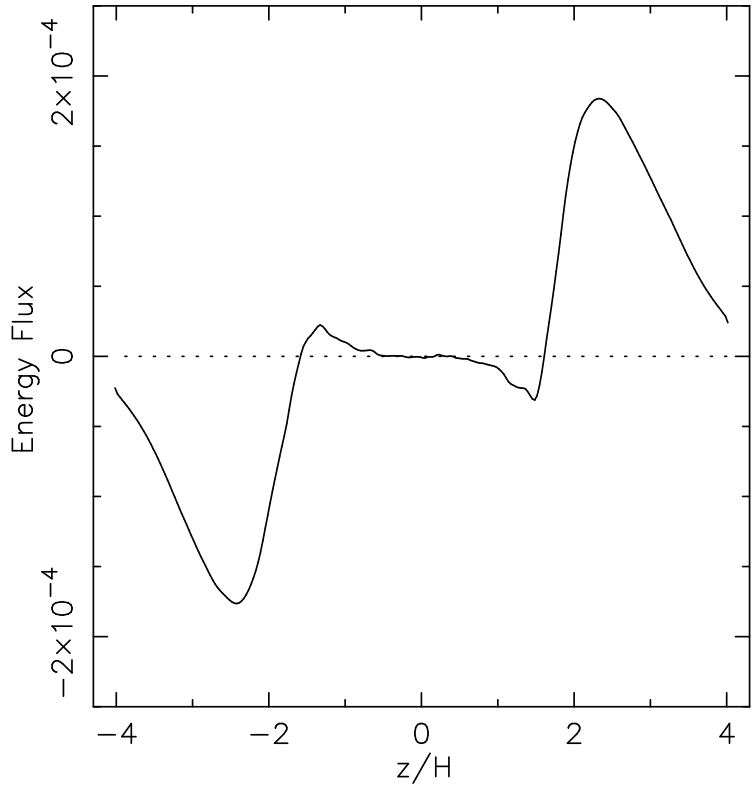}
\caption{Time-averaged Poynting associated magnetic tension, $-B_z v_{\perp} 
B_{\perp}$, flux of magnetic tension. }
\label{fig:efx}       
\end{figure*}

The disk winds are driven by Poynting flux; both magnetic pressures and 
tension almost equally contribute. The magnetic tension term, $-B_z v_{\perp} 
B_{\perp}$ ($\perp$ denotes $x$ and $y$) shows an interesting distribution, 
which is shown in Figure \ref{fig:efx}. 
In the surface regions, the absolute values decrease 
with increasing $z$, because the Poynting flux is converted to the kinetic 
energy of the disk winds; this is natural behavior. However, at $|z|\approx 
\pm 1.5H$, the solid line crosses the 0 line. This indicates that the Poynting 
flux of the tension force directs toward both surface and midplane from here.
This is a result of the breakups of large scale channel flows at 
$|z|\approx \pm 1.5H$ \citep{si09}.    

The Poynting flux associated with \Alfven waves is expressed as the same 
shape, $-B_z v_{\perp} B_{\perp}$, because they are propagating by tension 
restoring force. The properties of the Poynting flux of the magnetic 
tension shown in Figure \ref{fig:efx} are different from the \Alfven waves 
discussed in the solar and stellar winds, because $\delta B_{\perp} 
\gg B_z$ in the disk wind situation. The Poynting flux here is 
mainly by large-scale channel flows in mid altitude regions ($z\sim 2H$) 
and is associated with a U-shape magnetic field by Parker instability 
in the surface regions ($|z|\gtrsim 3H$: Suzuki et al.2010).

\section{Summary}
We have introduced the results of our simulation studies on \Alfven 
wave-driven winds. In the section \ref{sec:slw} we summarized the solar 
wind simulations presented in SI05 and SI06, 
emphasizing the reflection of \Alfven waves. In the standard run, 15\% of 
the initial wave energy flux from the photosphere can transmit into 
the corona, which is sufficient for the heating and acceleration of the 
solar winds. The \Alfven waves nonlinearly generate compressive waves 
which eventually steepen and heat up the ambient gas by the shock heating. 
In another way of the explanation, the outgoing \Alfven waves dissipate 
by parametric decay to incoming \Alfven waves and outgoing compressive 
waves. This process is very efficient in the density stratified 
atmosphere. 

The structure of the solar winds shows very sensitive dependence on 
the input wave amplitude by the positive feedback mechanism: When increasing 
the amplitude, the coronal density becomes larger and the pressure scale 
height becomes longer because of the larger heating. The larger density 
enhances the wave dissipation because the wave nonlinearly is larger 
owing to the smaller \Alfven speed ($\propto 1/\sqrt{\rho}$). The longer 
pressure scale height reduces the wave reflection. Therefore, the heating 
is enhanced when the surface amplitude slightly increases.  
This shows that the structure of solar wind is largely altered on a small 
change of the input energy through the reflection and the nonlinear dissipation 
of the \Alfven waves. 

We introduced recent HINODE observations of \Alfvenic oscillations. 
The obtained data by \citet{ft09} can be explained by the sum of upgoing 
\Alfven waves and the comparable amount of the reflected component. 
They can estimate the leakage Poynting flux of the \Alfvenic oscillations to 
the upper location from the phase correlation of the perturbing magnetic 
field and the velocity.  
They analyzed one case and found that the sufficient energy to heat the corona
and solar wind is going upward.    
These observed properties are quite consistent with what we get in the  
simulations.

We discussed the roles of \Alfven waves in the context of stellar 
evolution. Before reaching AGB phase when dusts are formed in cool atmosphere, 
it is expected that, similarly to the solar wind, the \Alfven waves play a 
major role in driving the stellar winds.  
By using the surface fluctuation strength estimated from the convective flux, 
we performed the MHD simulations. 
When a star evolves to RGB phase, the hot corona suddenly disappears and 
the atmosphere streams out as the cool chromospheric winds in which magnetized 
hot bubbles intermittently float. The main reason of the disappearance of 
the hot corona is the gravity effect; the atmosphere flows out before heated 
up owing to the weak gravity confinement. In addition, the thermal instability 
also plays a role in the sudden drop of the temperature. 
In evolved stars, the \Alfven waves do not suffer reflection because the 
density decreases more slowly and the change of \Alfven speeds become more 
gradual. This is a part of the reason why the mass loss rate jumps up 
with stellar evolution.

We also introduced the accretion disk winds by MHD turbulence amplified 
by MRI. The disk winds are driven by Poynting flux when the magnetic energy 
dominates the gas energy. In the disk wind regions the plasma $\beta$ value 
is kept slightly below unity by the redistribution of the 
density structure, which is qualitatively similar to what we got in the 
simulation of the solar and stellar winds. 
On the other hands, the disk winds are intermittent mainly because of 
the breakups of large scale channel flows. 
In the simulations, we only model the onset region of the disk winds. 
In further upper region with magnetically dominated condition, \Alfven waves 
might play a role in accerating the winds.

\begin{acknowledgements}
This work was supported in part by Grants-in-Aid for 
 Scientific Research from the MEXT of Japan 
 (19015004 and 20740100) and Inamori Foundation. 
\end{acknowledgements}

\bibliographystyle{aps-nameyear}      
\bibliography{example}   
\nocite{*}


\end{document}